# Cs diffusion in SiC high-energy grain boundaries

Hyunseok Ko, Izabela Szlufarska, Dane Morgan[*]

*Department of Material Science and Engineering, University of Wisconsin-Madison, WI 53706, USA*

## ABSTRACT

Cesium (Cs) is a radioactive fission product whose release is of concern for Tristructural-Isotropic (TRISO) fuel particles. In this work, Cs diffusion through high energy grain boundaries (HEGBs) of cubic-SiC is studied using an *ab-initio* based kinetic Monte Carlo (kMC) model. The HEGB environment was modeled as an amorphous SiC (a-SiC), and Cs defect energies were calculated using density functional theory (DFT). From defect energies, it was suggested that the fastest diffusion mechanism as Cs interstitial in an amorphous SiC. The diffusion of Cs interstitial was simulated using a kMC, based on the site and transition state energies sampled from the DFT. The Cs HEGB diffusion exhibited an Arrhenius type diffusion in the range of 1200-1600°C. The comparison between HEGB results and the other studies suggests not only that the GB diffusion dominates the bulk diffusion, but also that the HEGB is one of the fastest grain boundary paths for the Cs diffusion. The diffusion coefficients in HEGB are clearly a few orders of magnitude lower than the reported diffusion coefficients from in- and out-of- pile samples, suggesting that other contributions are responsible, such as a radiation enhanced diffusion.

*Corresponding Author. Tel:+1-608-265-5879 ; Email: ddmorgan@wisc.edu (Dane Morgan).

# 1 Introduction

One of the fuel type proposed for the next generation high-temperature nuclear gas reactors is Tristructural-Isotropic (TRISO)-coated fuel particle [1]. Recent TRISO fuels are designed to have a fuel kernel, typically $UO_2$ but can also be UC or UCO, surrounded by four successive layers consisting of C buffer, inner pyrocarbon, silicon carbide (SiC), and outer pyrocarbon [1]. The SiC layer mainly consists of the cubic 3C-SiC polytype, with a thickness of ~35 µm. The primary functional of this SiC layer is to provide a structural stability and to be a barrier for fission products (FPs). While this layer in the modern fuels effectively retains most of the FPs under both operating and accident conditions [2], some undesirable metallic FPs (e.g. Ag, Cs, and Sr) are historically reported to be released, particularly under accident conditions. Among these FPs, the release of the Cs isotopes has been long studied due to the large quantity of its production, and the relatively long radioactive half-life ($^{137}$Cs half-life is 30.2 years). Such release of Cs has significant safety and design implications.

To address issues associated with the Cs release from TRISO particles, it is important to first understand the transport mechanisms through the SiC that is the primary barrier in fuel particles. For past few decades, studies have been conducted with experimental or computational approaches to elucidate the diffusional rate of Cs and its release mechanism. The previously reported diffusion prefactors and activation energies for Cs are compared in Table 1, and these studies can be put into three categories, which will be briefly described in the following.

The first category of diffusion studies is based on integral release (I.R.) measurements. In these measurements, the effective diffusion coefficients are determined by fitting a simple diffusion model to the overall integrated FPs release data obtained from a batch of TRISO fuel particles [3,4]. While a few in-reactor data are available [5,6], most of these experiments measured the release of Cs from TRISO particles that were pre-irradiated then post-annealed at high temperatures. The effective diffusion coefficient through SiC is estimated by integrating with models for the Fickian diffusion through the other coatings in TRISO, and by adjusting the diffusion coefficients in SiC to match the observed FP release [7]. An interesting observation from combined I.R. measurements in Table 1 is that the activation energies are in the order of 1-2 eV for low-temperature annealing (i.e. less than about 1400 ˚C), while they are in the order of 4-5 eV for annealing at higher temperatures (above 1500 ˚C). The difference in activation energies suggests that there are two mechanisms which become dominate at different temperature regimes.

The second category of diffusion studies is based on ion-implantation (I.I.) measurements, where the diffusion of Cs in single crystalline (sc-SiC) 6H-SiC [8,9] and polycrystalline (pc-SiC) chemical vapor deposition (CVD)-SiC [8,9] were investigated using an ion implantation, followed by heating tests. These surrogate attempts can help understanding the release mechanism of FPs, however this method is known to create several complications. For instance, in I.I. studies the diffusion of Ag may be affected by the introduction of trapping sites from the implantation damage [10] and the change in diffusion mechanisms due to the higher concentrations of FPs than the concentrations in TRISO under operation [11]. Recently a novel design for the I.I. experiment was proposed [12] to avoid these complications, by introducing FPs into pyrocarbon adjacent to SiC without causing implantation damage in the SiC and at concentrations that would be relevant to TRISO fuel.

The third category of diffusion studies is based on computational studies, which have been carried out primarily to develop an atomistic level understanding of Cs diffusion. An *ab initio* study by Shrader *et al.* [13] showed that the Cs thermal diffusion in bulk is slow ($\sim 10^{-32}$-$10^{-25}$ m$^2$s$^{-1}$ at 1200-1600 ˚C), and these low bulk diffusion rates cannot account for the measurements in I.R.. However, Shrader *et al.* [13] found the activation energy for Cs bulk diffusion agrees with the I.R. measurements of temperature above ~1500 ˚C, which suggests Cs release occurs via bulk diffusion when irradiated. Recently, Rabone *et al.* [14] investigated Cs diffusion in $\sum 5$ coincide site lattice GB ($\sum 5$-GB) using an *ab initio* molecular dynamics at 1227 ˚C. This study showed that the Cs diffusion in $\sum 5$-GB is 3-4 orders of magnitude faster than the bulk diffusion.

The previous I.R. studies suggest that there are two diffusion mechanisms involved in Cs diffusion [15]. One hypothesizes to explain these trends is that grain boundary (GB) diffusion dominates at low temperature due to having an expected lower activation energy, while bulk diffusion dominates at a higher temperature where the higher temperature makes the migration energy less critical, and the larger accessible region for transportation allows the bulk diffusion to dominate. Another hypothesis, proposed by Malherbe [15], is that there is de-trapping of Cs atoms from irradiation-induced defects in the bulk at high temperature, and these de-trapped Cs then diffuse via a volume diffusion. However, the large variations within the I.R. measurements and the difference in their experimental condition (e.g. thermal exposure time) make it hard to determine which transport mechanism dominates, and how quickly Cs releases. Furthermore, it remains unclear if there is any radiation effect on the release of Cs and if yes, a quantitative assessment is necessary to determine how significant the effect is.

As mentioned above, the GB diffusion is postulated as a dominant mechanism in low temperatures, but a critical missing part of the present understanding is the diffusion mechanisms and diffusion rates in GBs. Among GBs, the high-energy GBs (HEGBs) are expected to play a significant role in the FPs transportation in pc-SiC, for following reasons. The HEGBs are often highly disordered structures and represent > 40% of GBs in TRISO prototype materials with grain sizes of 0.5-1.5 $\mu$m[16,17]. The high fraction of HEGBs allow them to provide a percolating path for FPs transportation, and HEGBs are one of two GB types that are present in a high enough concentration to form a percolating path (the other GB type is $\sum 3$-GB) [18]. Also, it is often found that disordered (amorphous) materials, which are structurally similar to HEGBs [19,20], provide a faster transportation pathway for extrinsic defects compared to crystalline materials [21-23]. Furthermore, our recent study has suggested that Ag diffusion via HEGB is the dominant path for diffusion in unirradiated SiC [24]. Therefore, we expect HEGBs are the most likely GB type to dominate Cs transport, and the HEGB Cs diffusion coefficient to be faster than those in other GBs or bulk. In this work, we used an *ab-initio* based stochastic modeling approach to predict the Cs diffusion in HEGBs. Based on the diffusion coefficient from the model, we will determine under what conditions (e.g. temperature), if any, HEGB *D* is a dominant pathway in SiC, and responsible for the release of Cs.

## 2 Methods

The approach taken in this work follows that used for a similar study on Ag in SiC from Ko *et al.* [24]. We briefly repeat the essential details here, with relevant changes for treating Cs in place of Ag.

### 2.1 *Ab-initio calculations*

Modeling of the HEBGs, often called disordered GBs, has been a great challenge in covalent materials [25]. In this work we follow the approach of Ref. [24] and model the HEGB structure as an amorphous SiC (a-SiC) because (i) the local environments in HEGBs of covalent materials are known to be similar to amorphous phases [19,20], and (ii) a bulk a-SiC is computationally more tractable than modeling a full GB structure. A simulation cell is used that is approximately cubic and contains 128 atoms (64 Silicon, and 64 Carbon). The cell was trimmed from a bulk a-SiC, which was prepared with the melt-quench method [26] using a Tersoff potential [27]. This supercell was then fully relaxed (constant zero pressure and zero K temperature) using density functional theory with the periodic boundary conditions. The relaxed supercell had cell vectors of [11.50, -0.12, 0.04; -0.12, 11.45, -0.10; 0.04, -0.08, 11.51] (non-cubic). The supercell used in this study is shown in Figure A. 1 of Appendices. The density of SiC in this supercell, after fully relaxed with DFT, was 2.81 g/cm$^3$, which was comparable value with the simulated a-SiC density 3.057 and 2.896 g/cm$^3$ [26] (by Tersoff potential), and was lower than the sc-SiC density of 3.18 g/cm$^3$ (calculated with DFT), as expected.

Ab initio density functional theory (DFT) calculations were performed using the Vienna Ab-Initio Simulation Package (VASP) [28-31]. The VASP calculations were performed using the projector-augmented plane-wave (PAW) method [32,33] and the generalized gradient approximation (GGA) exchange-correlation potentials, parameterized by Perdew, Burke, and Ernzerhof (PBE) [34]. The PAW file electronic configurations used are: 3s2 3p2 for Si, 2s2 sp2 for C, and 5s2 5p6 6s1 for Cs. These pseudopotentials have been widely used in SiC studies [35-37] and Cs studies [13], therefore no benchmarks are presented in this study. The energy cut-off was set as 450 eV, and a single Γ-point *k*-point mesh was used to sample reciprocal space. The plane wave energy cutoff and the *k*-point mesh density were separately tested for eight cases and were also converged to give a total energy within 180 meV (with minimum of 32 and maximum of 175 meV) per Cs defect compared to 3×3×3 *k*-point mesh, where this larger mesh is expected to be well-converged.

To obtain diffusion pathways and the corresponding energy barriers, the climbing image nudged elastic band (CI-NEB) method was employed [38]. For CI-NEB calculations three images were used and images were linearly interpolated between minima. Following the approach in Ref. [24], in the cases of images where the Cs atom is too close to Si or C atoms, their positions were adjusted to be separated by 75% of their covalent bond length (2.7 Å for Cs-Si, and 2.4 Å for Cs-C [39]). The *k*-point mesh test on eight barriers showed that the error in the energy barrier with respect to *k*-points for a Γ-point vs. 2×2×2 *k*-point mesh is within an acceptable error range of 220 meV/Cs (with minimum of 53 and maximum of 216 meV), which corresponds to about a factor of 5× error at 1500K.

For the defect formation energy ($\Delta E_f$) of defects, we used the following expression [40]:

$$\Delta E_f = E_{def} - E_{undef} + \sum_I \Delta n_I \mu_I. \quad (1)$$

where $E_{def}$ and $E_{undef}$ are energies of the defected and the undefected cell, $\Delta n_I$ is the change in the number of the atomic species *I* in the defected cell from the number of same species in the undefected cell. The $\mu_I$ is the chemical potential of atomic species *I* relative to its reference states, which are taken as the bulk Si and C VASP energies (which are in turn referenced to the appropriate atomic species energies given as defaults in the pseudopotential files) in their groundstate structures (diamond lattice for Si and graphite for C). These values are $E_{Si}$=-5.431, and $E_C$=-9.186 eV/atom. Throughout this study, the Si-rich condition chemical potentials are used for silicon and carbon ($\mu_{Si}$ = -5.431 and $\mu_C$ = -9.619 eV) for consistency. It should be noted that the chemical potential for each element of a binary system are not unique[41-43] and may vary within a range between Si-rich or C-rich conditions. Under C-rich conditions, C based defects will increase their formation energy in the positive direction (toward being less stable) by the calculated heat of formation of SiC (0.44 eV). However, this will not affect our results for Cs diffusion via interstitial mechanism, since interstitial is the most stable form of Cs defect regardless of the chemical potential choice. The *ab-initio* formation energy of bulk solid phase Cs metal ($\mu_{Cs}$ = -0.856 eV) was set to be the chemical potential for Cs. We found that the interstitial diffusion is expected to be the dominant mechanism in the HEGB and it will be the focus in this study (further discussion in Sec. 3.1. and Table 2). The result was also found to be independent of the chemical potentials chosen for Si and C.

It should be noted that no charged supercells or explicitly charged defects were considered in the present study. We believe this approximation is reasonable as the neutral state for Cs interstitials is the stable charge state for Cs in *n*-type crystalline SiC, as predicted by Shrader, *et al* [13]. Also, the effects of the association of Cs defects with intrinsic defects and the interaction between Cs defects had not been considered. The results should be interpreted within these limitations and are an indication of what the behavior would be if only non-interacting defects are present in the material.

To model Cs interstitial ($Cs_I$) diffusion in an a-SiC, the energy landscape of Cs interstitials was first investigated. In an amorphous material, locating extrinsic interstitial defect sites has many complications due to the lack of long-range order. In this paper, we used a simple gridding method, in which we grid the entire supercell with a fine uniform grid and relax a Cs interstitial at every grid point in the cell, to obtain a complete list of possible $Cs_I$ sites. This method for finding sites is more computationally expensive but more comprehensive when compared to approaches often taken in previous studies, such as geometry guided guesses [44,45] and free-volume based search [46]. In this method, the entire supercell is gridded by a 1 Å grid along each axis, and Cs interstitials are placed at every grid points then relaxed to search all existing sites. The details of gridding method, including the choice of grid size and an assessment of its ability to identify all interstitial sites, can be found in our previous work on Ag diffusion in HEGB [24].

*2.2   Kinetic Monte Carlo model*

The modeling of diffusion in the amorphous systems has been a challenging problem in many applications. One of the most pragmatic and simple method is by molecular dynamics with interatomic potentials, however, potentials for such complex systems (e.g. FPs in SiC) are not developed. There are other plausible methods such as *ab-initio* molecular dynamics [47] or potential energy surface sampling approach (e.g. kinetic activation relaxation [48,49], and the autonomous basin climbing [50]) can be used but the computational demands of such calculations are often prohibitive with full *ab-initio* methods. Therefore, in this work, we use an approach based on kinetic Monte Carlo (kMC). This section describes the methods used in the KMC model and the results of using these approaches are given in Sec. 3.2.

An essential input for the KMC model is the energy landscape for hops. In our previous study of Ag diffusion in HEGB [24], we modeled an effective medium (virtual lattice) that mapped the effective energy landscape of Ag interstitials. However, to avoid additional complexity and uncertainties introduced by the effective medium approach it was not used in this study. Instead, the diffusion was directly modeled by applying periodic boundary conditions to the simulation cell where the Cs interstitial sites were identified by the DFT calculations. A similar approach in a finite cell with periodic boundary conditions successfully modeled the Li diffusion in amorphous (a-) materials, e.g., a-Si [51], a-$Al_2O_3$ [52], and a-$AlF_3$ [52], where they generally showed good agreements with experimental diffusion rates when the comparison was possible.

Overall, the 55 interstitial sites have 3.4 neighbors within 4.0 Å, which result a large number of hopping barriers to determine. To reduce the number of calculations barriers were only calculated until the diffusion coefficient ($D$) showed convergence with respect to the barrier sampling. Here we describe the approach used in detail. The $Cs_I$ sites were identified first, then migration barriers between these sites were calculated. To reduce the number of hops that needed to be considered, it was assumed that the diffusion was likely dominated by the more stable $Cs_I$ sites. With this assumption, the most relevant migrations barriers were determined as follows. A set of sites was created by taking the *n* most stable sites. Then all migration barriers between all the ($n!/2$) pairs were determined with DFT for any pair of sites within a cutoff distance of 4.0 Å. This was the longest distance over which no intermediate state was found from preliminary test calculations of randomly sampled 25 barriers. This procedure was done first for *n*=1 (i.e., the most stable site), then *n*=2 (i.e., the 2 most stable sites), and so on.  At each *n,* the migration barriers determined between those *n* sites were used in the kinetic Monte Carlo simulation to evaluate the Cs diffusion coefficients on the network of those *n* Cs interstitial sites.  This process was done for increasing *n* until the diffusion coefficient was converged, i.e., stopped changing as *n* increased.

As a further test of convergence for a given *n*, migration barriers for all other $Cs_I$ sites were estimated and are also included in the model, to see if the estimate for *D* changed. The approach to estimating the values not calculated with DFT made use of the kinetically resolved activation (KRA) formalism [53]. First we define some notation. We refer to the network of *n* interstitial sites and their calculated barriers as subsystem $S_n$. We refer to $S_n$ with additional barriers estimated by the KRA formalism ($E_{KRA}$) for all remaining interstitial sites as $S_n^+$. We refer to the diffusion coefficient calculated on the subsystem $S_n$ or $S_n^+$ by $D(S_n)$ or $D(S_n^+)$, respectively. In a simple KRA approach unknown migration energies between sites X and Y are estimated from the energies at sites X and Y by setting the transition state energy between them, $E_{TS}(X,Y)$, to:

$$E_{TS}(X,Y) = \frac{(E_X + E_Y)}{2} + E_{KRA};$$
$$\text{if } E_{TS} < \max(E_X, E_Y), \text{ then } E_{TS} = \max(E_X, E_Y). \qquad (2)$$

where $E_X$ and $E_Y$ are site energies of $X^{th}$ and $Y^{th}$ stable $Cs_I$ and $E_{KRA}$ is approximated as a constant. For $E_{KRA}$ in the system $S_n^+$, we have used the $E_{KRA}$ value of zero. This approximation sets $E_{TS}(X,Y) = \max(E_X, E_Y)$, which leads to barriers of $|E_X - E_Y|$ for hops that increase energy and zero for hops that decrease energy. This approach provides an estimated upper bound on the $D$ one might get using DFT values for all the interstitial sites. When $D(S_n)$ stops changing with $n$ for three consecutive $n$ values and $D(S_n) = D(S_n^+)$ for all these $n$ values, we conclude that the system size $n$ is sufficiently large to include all essential interstitial sites along the diffusion path, and thus, the $D$ is converged. One problem with this approach is that for a few sites the very low barriers (which occur when $E_X \approx E_Y$, as $E_{KRA}=0$ then leads to a near zero barrier) caused the Cs to become trapped hopping between just two sites during the KMC calculations. For these cases that led to trapping (there were four for $n=23$), we calculated and used the DFT value of the barrier, which was always much higher than the lower-bound estimate, and therefore removed the trapping.

The kinetic Monte Carlo (kMC) method was employed with the Bortz-Kalos-Liebowitz [54] algorithm. The hopping rates ($\Gamma$) for Cs atoms were given by the transition state theory as $\Gamma = \omega \cdot e^{(-E_m/k_BT)}$, where $\omega$ is attempt frequency and $E_m$ is the migration barrier of the hop. The attempt frequency ($\omega$) can be determined as $\omega = \prod_{3N} \nu^{equil} / \prod_{3N-1} \nu^{saddle}$, where $\nu^{equil}$ and $\nu^{saddle}$ are the vibrational frequencies at the equilibrium and saddle point, and $N$ is a number of atoms in an adequately large region around the diffusion path [55]. We took $N$ to be all atoms within a 4 Å cutoff of the Cs in the transition state of the hop being considered, which was typically 12–16 atoms. Tests for longer cutoff distances and larger $N$ gave change in $\omega$ of less than 5%. From *ab-initio* calculations for 5 test cases, we found that the attempt frequencies typically range from $10^{12}$ to $10^{13}$ s$^{-1}$, with an average of $3.8 \times 10^{12}$ s$^{-1}$. This average value was then used for all hops.

The diffusion coefficients were determined by the Einstein relation, $D = <r^2(t)>/2dt$, from the calculated mean square displacement as a function of time [56], where $d$ is the dimensionality of the system (here $d = 3$), $t$ is time, and $<r^2(t)>$ is the mean square displacement of Cs as a function of time. We evaluated the average using the multiple time origin method [57]. The details of the kMC can be found elsewhere [24]. Simulation at each $n$ was typically performed for $5 \times 10^9$ kMC steps to obtain a well-converged diffusion coefficient.

## 3 Results

### 3.1 *Ab-initio calculations*

To understand the transportation behavior of Cs in a-SiC, it is important to determine which of the defect types are stable and will contribute to the diffusion. The formation energies ($E_f$) of the point defects (vacancies, Cs substitutionals on Si and C lattices, and Cs interstitials) were investigated. Each of these defects was sampled over a wide range of local environments of a-SiC. In Table 2, the average and standard deviation of $E_f$ values over many defect sites in the a-SiC are shown to represent the spread in values. The formation energies of these defects in the sc-SiC are also presented for a comparison. The sampled set includes 15 substitutionals and 55 interstitials, where this latter set is quite comprehensive as they were determined to be the dominant mechanism and studied in detail (see Sec. 3.1). All the defect types in a-SiC, on average, had lower $E_f$ than those in the sc-SiC. Particularly, Cs interstitials in the a-SiC showed a dramatic difference compared to $E_f$ in sc-SiC. For all identified sites, the mean $E_f$ of Cs interstitials in a-SiC was lower by > 13 eV, compared to the most stable Cs interstitial (23.46 eV) in sc-SiC (6–fold coordination for Cs, with three carbon and three silicon atoms). Some of the Cs interstitial formation energies were near zero or negative, i.e. the Cs interstitials are more stable in the a-SiC than in a bulk metallic Cs. This result suggests that Cs will segregate strongly to disordered HEGBs, consistent with what was found in the previous GB modeling study [18].

Under the dilute Cs condition, the diffusion can take place via either a vacancy-mediated substitutional diffusion or some form of interstitial diffusion. The vacancy-mediated diffusion mechanism, however, requires a migration of a vacancy [58]. The barriers to migrating a vacancy around Cs in a-SiC were found as 2.6 – 4.5 eV, and the barriers to migrating a vacancy in a undefected a-SiC were 3.0 – 4.5 eV (consistent with the previous study [24]) from tests on five barriers for each case. These values suggested that the vacancy mediated Cs diffusion is expected to be relatively slow compared to interstitial migration, which can avoid moving Si and C atoms. Therefore, in this study, we only considered the interstitial diffusion mechanism for Cs in the a-SiC. Our results below predicted a migration barrier of 2.41 eV through the interstitial mechanism. As this value is smaller than the ranges above, it strongly supports our hypothesis that interstitial diffusion is dominant for Cs in a-SiC. No interstitialcy (or kickout) mechanisms [59] were considered.

By applying gridding to the a-SiC sample, we investigated Cs interstitials on 12 ×12 ×12 grid points. The interstitial sites that were both energetically and geometrically close to each other were grouped and represented by the lowest energy site in the group. We have referred this grouping as a hierarchical clustering approach and the details are available elsewhere [24]. After hierarchical clustering, a total of 55 Cs interstitial sites were identified. In Figure 1, the distribution of formation energies is shown. The formation energies of Cs interstitials in a-SiC had an average value of 0.41 eV and this value is ~23.1 eV more stable than the most stable Cs interstitial in sc-SiC, as can be seen in Table 2. The values ranged from -2.68 to 8.15 eV, with a standard deviation of 3.40 eV. The Cs, which is an impurity with a relatively larger size than those studied in the similar amorphous systems [46,47,51,60-63], turned out to be highly stable as interstitials in the a-SiC. The high stability of Cs defects vs. bulk is possibly due to the large relaxations available to the a-SiC system, analogous to the observation of Ag defects in the HEGB [24], although the detailed mechanism of stabilizing Cs interstitials in a-SiC was not explicitly investigated here.

After identifying interstitial sites, the migration barriers were calculated for the Cs. Initially we randomly sampled 25 $E_m$ and associated kinetically resolved activation barriers ($E_{KRA}$), which can represent migration barriers independently of the initial and final state energies (details are presented in our previous study, Ref. [24]). As it can be seen from the Figure 2, more than 40% of $E_{KRA}$ were found less than 1 eV, suggesting a fast-interstitial diffusion may be possible. To obtain the complete set of barriers and hopping paths we have sampled migration barriers by increasing sampling size ($n$) starting from sites with the lowest energy, as explained in Sec. 2.2 and with results given in Sec. 3.2.

### 3.2 Kinetic Monte Carlo modeling

Ideally, all the migration between 55 Cs interstitial sites can be sampled to obtain the effective diffusion coefficient in HEGB ($D_{HEGB}$), but this would be computationally intractable. As discussed in Sec. 2.2, we have assumed Cs visits only the more stable sites during its diffusion. Therefore, the kMC simulations were preformed at set of sites of size $n$, where $n$ was increased to include less stable sites until $D$ showed convergence. Furthermore, following the approaches discussed in Sec. 2.2, we have calculated $D$ for system of sites and barriers associated with the $n$ points ($S_n$) and a system of all points where additional barriers are calculated by the KRA method ($S_n^+$). The resulting $D(S_n)$ and $D(S_n^+)$ from two methods are shown in Figure 3. For $n \leq 8$, there was no diffusion across the boundary of the simulation cell for both $S_n$ (no diffusion path with small $n$) and $S_n^+$ (Cs is trapped in few $n$). When a faster diffusion path is added ($S_{n=9}$, $S_{n=12}$, $S_{n=18}$, $S_{n=9}^+$, and $S_{n=18}^+$), a steep increase in $D$ is observed, as shown in Figure 3. For $9 \leq n \leq 20$, we have $D(S_n^+) > D(S_n)$. This is because the sites that were not included in $S_n$ system, i.e. the (55-$n$) sites, were playing a role in Cs diffusion as they generally offer hops with low $E_m$s (they are estimated with $E_{KRA} = 0$ eV, which provides very low barriers). For $n \geq 21$, the convergence of $D$ with $n$, and the fact that $D(S_n) = D(S_n^+)$ support that $n = 23$ is large enough to model $D$ for the full set of 55 Cs interstitial sites. The diffusion coefficients reported in the following were calculated in the $S_{n=23}$.

The diffusion coefficients using $n = 23$ were simulated for 1200–1600°C, and are summarized in Figure 4. In this temperature range, the $D_{HEGB}$ values ranged from $10^{-19}$–$10^{-17}$ m$^2$s$^{-1}$, which is 4–6 orders of magnitude higher than the bulk diffusion coefficient ($D_{bulk}$) [13]. The HEGB Cs diffusion exhibited an Arrhenius type diffusion as shown in Eq.(3),

$$D_{HEGB} = D_0 \exp^{(-E_A/k_b T)}. \quad (3)$$

where $D_0$ is the diffusion prefactor, $E_A$ is the effective activation energy for the interstitial diffusion, and $k_b$ is the Boltzmann constant. This Arrhenius form fitted to the KMC simulations yielded $D_0$ and $E_A$ values of $(3.75 \pm 3.15) \times 10^{-11}$ m$^2$s$^{-1}$ and $2.41 \pm 0.26$ eV, respectively. It is interesting to note that for a simple ideal interstitial diffusion model we expect the prefactor, $D_0$(ideal), to be $D_0$(ideal) $= gl^2\omega$, where $g$ is a geometric factor of order one, $l$ is the hop length, and $\omega$ is the attempt frequency. Using the attempt frequency given above for our model and a hop length of 3 Å (similar to our average hop distance) yields an estimate of $D_0$(ideal) =

(3.4)×10$^{-7}$ m²s$^{-1}$, about 10$^4$ times larger than our fitted value. A similar suppression of the $D_0$ value vs. a simple model was found for Ag in SiC in Ref. [24]. While the $D_0$ value can be strongly affected by small deviations from Arrhenius behavior when fitting in only a small range at high temperatures, this result also suggests that the Cs diffusion exhibits significant reduction in diffusion relative to a simple model, presumably due the disordered energy landscape through which it must move.

It should be noted that our Cs HEGB diffusion is modeled by a limited size simulation cell, and this cell is likely not fully capable of representing the a-SiC nature. For this reason, $D_{HEGB}$ is expected to have some uncertainty originating from the sampling. This uncertainty had been estimated in the previous work on the Ag diffusion in a-SiC [24]. By taking the uncertainty estimated in Ref. [24], we expect the $D_{HEGB}$ to have an uncertainty of approximately a factor of 10 (up to 100 at the most) due to limited structural sampling.

## 4 Discussion

The HEGBs are expected to be the dominant diffusion path for Cs not only because of the diffusion rate, but also because the highest fraction of GBs are HEGBs and because Cs clearly has strong segregation tendencies to the HEGB. Here we discuss each of these properties, their relation to net diffusion in pc-SiC, and the comparison to other Cs release studies.

As mentioned in the Introduction, for HEGBs to be a dominant Cs release path, they must enable a connected (percolating) pathway through the polycrystalline material. In pc-SiC, $\sum$3-GBs and random GBs (HEGBs) are the most abundant type of GBs [16,64] and the HEGBs constitute a majority, more than 40%, of GBs in CVD-SiC [17]. Among the GB types only HEGBs and $\sum$3-GBs, with their high fraction of all GBs, are expected to provide a percolation paths for FP diffusion in SiC [18].

If HEGBs form a percolating network, the diffusion rates in HEGBs must be compared to other possible paths in pc-SiC to determine which is fastest. The present work showed HEGBs can provide much faster diffusion paths compared to the bulk [13], suggesting that the GB diffusion dominates over bulk diffusion in most cases. Study on Cs diffusion in a low angle $\sum$5-GB by Rabone *et al.*[14] predicted a range of $D_{\sum 5\text{-}GB}$ to be 2–4 orders lower than $D_{HEGB}$ as shown in Figure 4. Unfortunately, studies on the GB diffusion are very limited in a number, and it is not possible to compare diffusion rates in a range of different GB types. In particular, the diffusion data in $\sum$3-GBs, which is also likely to form a percolating path, needs to be determined to conclude that the HEGB is indeed the fastest and dominant pathway.

Even though there is no $D_{\sum 3\text{-}GB}$ data available, we hypothesize HEGBs are likely to provide the fastest path among the GBs. This hypothesis is reasonable because the diffusion barriers of Cs in $\sum$3-GB are expected to be greater than those in HEGB. Assuming the diffusion prefactors in the Arrhenius equation are a similar order of magnitude for both $\sum$3-GB and HEGB, the effective activation barrier contributes dominantly to their diffusion coefficient. The effective activation barrier in HEGB is expected to be lowest, as the diffusion barriers are, in general, inversely proportional to the free volume of the system. This idea is supported by the Cs migration barrier of 4.65 eV in the $\sum$5-GB [14], whereas the effective barrier in the HEGBs was 2.41 eV. Also, in the similar study of Ag diffusion in unirradiated SiC, the $D_{\sum 3\text{-GB}}$ [18] was found to similar or lower than $D_{HEGB}$ [24]. Analogous trends are expected for Cs, perhaps favoring HEGBs even more, as Cs is much larger atom than Ag and therefore the movement of Cs will be sluggish in the more confined geometry of the $\sum$3-GB.

Assuming HEGBs are the dominant GB diffusion pathways, it is important to consider the effect of microstructural features on the effective diffusion we expect in pc-SiC. The net diffusion coefficient in pc-SiC ($D_{net}$) allows the comparison to the diffusion coefficients measured from both I.I. and I.R. in pc-SiC. Diffusion coefficients obtained in I.R. measurements are effective values that combine contributions from all transport mechanisms available to Cs. The $D_{net}$ is given by the modified Hart equation [65]:

$$D_{net} = [\delta s D_{GB} + (d - \delta)D_B] / (\delta s + d - \delta) . \qquad (4)$$

where $s$ is segregation factor, $d$ is grain size, and $\delta$ is GB width. For $D_{GB}$ we approximate it to be $D_{HEGB}$. The segregation factor ($s$) can be calculated as:

$$s = C_{GB}/C_{Bulk}. \qquad (5)$$

where $C_{GB}$ and $C_{Bulk}$ are equilibrium Cs concentrations at GB and bulk. We assume $C_{GB}$ is approximately equal to $C_{HEGB}$. The equilibrium concentrations can be calculated from formation energies of Cs defects in GB and bulk, as shown in Eq. (6):

$$C^{Cs} = \sum_i \frac{\exp(-E_f^i/k_B T)}{(1+\exp(-E_f^i/k_B T))} \times \rho_i . \qquad (6)$$

where $i$ is the type of defect, $\rho_i$ is the site density of defects of type $i$ per unit volume, and $E_f^i$ is the formation energy of the defect of type $i$. With this simple non-interacting model for the Cs solubility, we predict the solubility limit in the HEGB ($C_{HEGB}^{Cs}$) to be $2.25\times10^{26}$ m$^{-3}$ at 1200°C. The solubility limits in the sc-SiC can be calculated using values from Ref.[13], as $C_{bulk}^{Cs} = 1.9\times10^{-9}$ m$^{-3}$. From Eq.(5), the segregation factor is found as $s \sim 10^{35}$ at 1200°C, and $s \sim 10^{27}$ at 1600°C. Although there is no experimental evidence showing a Cs segregation to GB (while Ag atoms are detected in GBs and triple junctions [64]), these large values of $s$ clearly suggests a strong segregation of Cs to GBs (HEGBs).

The $D_{net}$ can be estimated from Eq.(4) by using $s$ values calculated from above, and using experimentally measured values of $d$ and $\delta$, which are taken as 1μm and 0.5nm, respectively. Due to high segregation factors at all relevant temperatures, $D_{net}$ becomes approximately equal to $D_{HEGB}$ (the difference is less than 0.001% of $D_{HEGB}$, at most). In other words, the GB diffusion governs the total diffusion. While it is hard to predict the real diffusion process in the radiated material, a simple diffusion kinetic model involving both bulk and GB (type A and B, Harrison's model [66]) is unlikely based on the analysis presented here.

The contribution of $D_{HEGB}$ to the net diffusion in pc-SiC can be determined by comparing this value to previously reported I.I. and I.R. measurements. For ion implantation study, Audren *et al*.[67] and Friedland, van der Berg, Hlatshwayo, Kuhudzai, Malherbe, Wendler and Wesch [8] have investigated Cs diffusion in the ion implanted SiC. In both studies, no Cs diffusion was observed below the certain temperature (around 1050 and 1200 °C, respectively), and Cs started to become mobile above those temperatures. Friedland, van der Berg, Hlatshwayo, Kuhudzai, Malherbe, Wendler and Wesch [8] postulated that Cs impurities are trapped by the implantation damage, thus no movement of Cs is observed. The implantation damage is a disadvantage in the ion implantation studies, while it is not a concern in the TRISO case. To resolve the issue, Dwaraknath *et al.* [12] have developed a novel design for I.I., where the FPs are implanted in a thin pyrocarbon layer which layer is in between two SiC layers. In this way, no direct ion implantation damage is done on the SiC layer. Dwaraknath *et al.* [68] showed that Cs diffusion indeed takes place at a temperature range of 900–1300 °C, if there is no implantation damage. The calculated $D_{GB}$ from Ref. [68] showed a good agreement with the $D_{HEGB}$ calculated in this work, giving measured values about an order of magnitude lower over a wide temperature range, as it can be seen in Figure 4. It is worth noting that a factor of 10 is within the uncertainty we expected due to the sampling from a finite cell. Furthermore, assuming an Arrhenius behavior, the activation energy from Dwaraknath *et al.* [68] is 2.6 $\pm$0.6 eV, which compares very well to the values predicted in this work of 2.41 $\pm$ 0.26. The agreement strongly supports that the HEGB is the dominant diffusion path for Cs in non-irradiated SiC. However, it is worth noting that the bulk diffusion coefficient determined by Dwaraknath *et al*. is orders of magnitude higher than that predicted from *ab-initio* methods by Schrader *et al*. [13], as well as having a much lower activation energy (1$\pm$ 0.1 eV from Dwaraknath *et al*. vs. 5.14 eV from Schrader, *et al*.). The origin of this discrepancy is still unclear and suggests a need for further study of possible bulk mechanisms for Cs transport.

The diffusion coefficients extracted from I.R. measurements ($D_{IR}$) are greater than the $D_{HEGB}$ at all temperature ranges, as it can be seen in Figure 4. Particularly, in the relevant temperature ranges of 1200–1600 °C, there are approximately 1–3 orders of magnitude difference between the predicted $D_{HEGB}$ and the $D_{IR}$. Compared to I.R. measurements, the ion implantation measurements by Dwaraknath *et al*. [68] also exhibit a low value, quite consistent with our calculations. The I.I. experiments and our calculations are under conditions quite different from I.R., but perhaps the most obvious difference is that I.R. measurements all involved irradiation. Therefore, we speculate that the predicted $D_{HEGB}$ is representative of the effective diffusivity in unirradiated pc-SiC, but that irradiation accelerates the Cs diffusion. In general, radiation enhanced diffusion (RED) has been widely observed in metallic nuclear materials and semiconductors [69-71]. More specifically, a recent study on Cs diffusion in SiC by Dwaraknath and Was [72] observed RED of Cs in Si$^{++}$ irradiated pc-SiC, and the enhancement was about 2 orders of magnitude in Cs diffusion, even when the diffusion was measured after the irradiation had stopped. At this point, it is not clear how the irradiation might enhance the Cs transport, and whether GB diffusion is the active mechanism, perhaps with RED, with irradiation. There is any no detailed determination of the evolution of HEGB structure in SiC under irradiation of which we are aware. However, given that the grains are stable after irradiation[73], we do not expect the HEGB to completely transform or be removed from the system. Nonetheless, as we conclude that irradiation will change the Cs diffusion, there may be structural changes in HEGBs which lead to these transport changes. Furthermore, there are other amorphous systems that show structural changes under irradiations[74]. Overall the present study supports that HEGB is the dominant pathway for Cs release in unirradiated SiC, and enhancement is expected due to the radiation.

## 5 Conclusion

In summary, Cs defects in a-SiC (used to represent a HEGB) were calculated with DFT calculations and Cs diffusion in HEGB was modeled using the kMC method. The interstitial was found as not only the most stable defect type, but also as the fastest diffusion mechanism for Cs in a-SiC. The low formation energies of Cs defects indicated a strong segregation to GB. The Cs interstitial diffusion coefficients were calculated for 1200–1600 °C, over which range they exhibited Arrhenius type diffusion. The diffusion prefactor and effective barrier values were calculated as $(3.75 \pm 3.15) \times 10^{-11}$ m$^2$s$^{-1}$ and $2.41 \pm 0.26$ eV, respectively. We predicted that $D_{HEGB}$ is 4–6 orders higher than $D_{bulk}$ in the operating temperature range. A good agreement with $D_{GB}$ from a recent ion implantation study suggested that the HEGB is the dominant pathway for Cs diffusion in unirradiated 3C-SiC. However, the discrepancy between $D_{HEGB}$ and $D$ from ion implantation studies with irradiation and integral release measurements suggested that radiation enhance diffusion may occur for Cs in SiC.

## 6 Acknowledgement

This material is based upon work supported by the U.S. Department of Energy office of Nuclear Energy's Nuclear Energy University Programs under grant No.12–2988. Computations in this work benefitted from the use of the Extreme Science and Engineering Discovery Environment (XSEDE), which is supported by National Science Foundation grant number OCI–1053575.

## 7 Appendixes

A. Amorphous SiC (a-SiC) structure

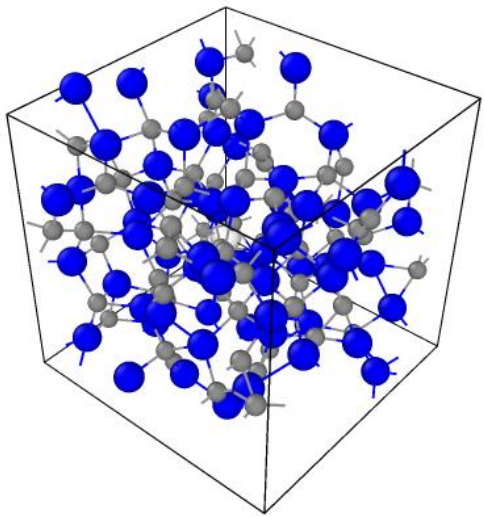

Figure A. 1. a-SiC simulation cell used in this study. The cell contains 128 atoms, 64 Si atoms and 64 C atoms.

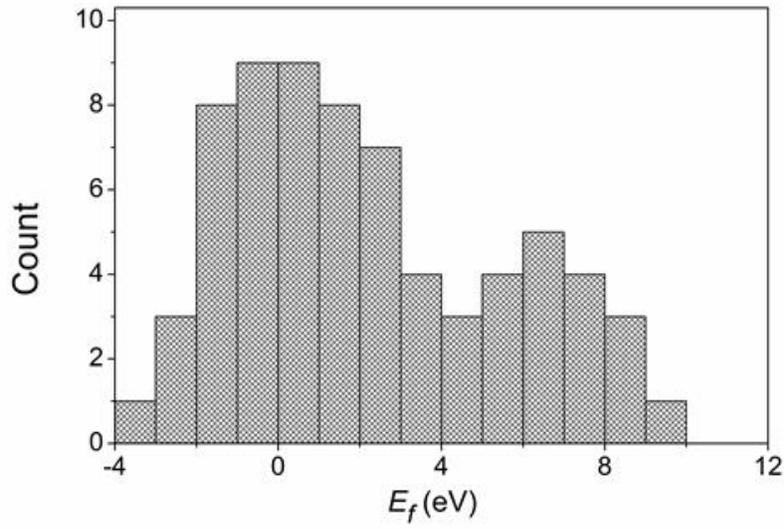

Figure 1. The distribution of formation energies ($E_f$) for identified 55 Cs interstitial sites in a-SiC.

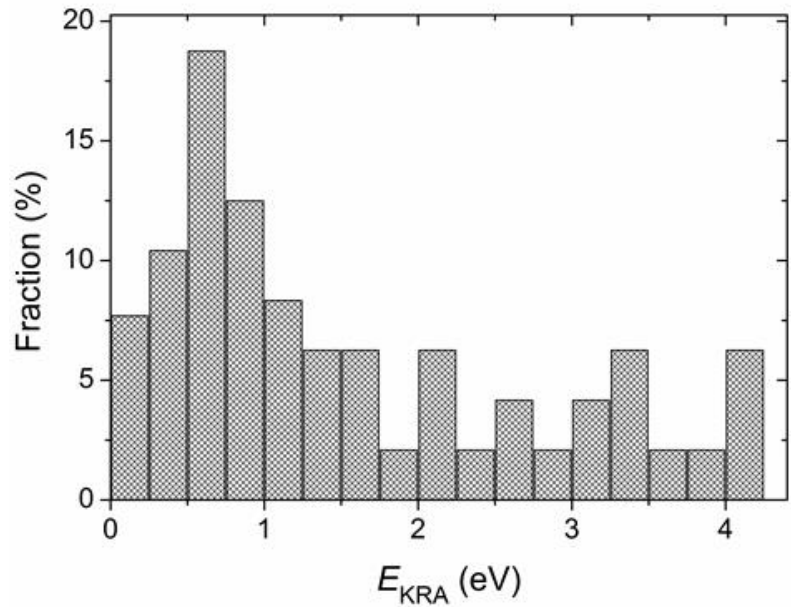

Figure 2. The distribution of kinetically resolved activation barrier for 96 migration between Cs sites (30 randomly sampled, and 66 sampled for most stable sites as described in Sec. 3.1).

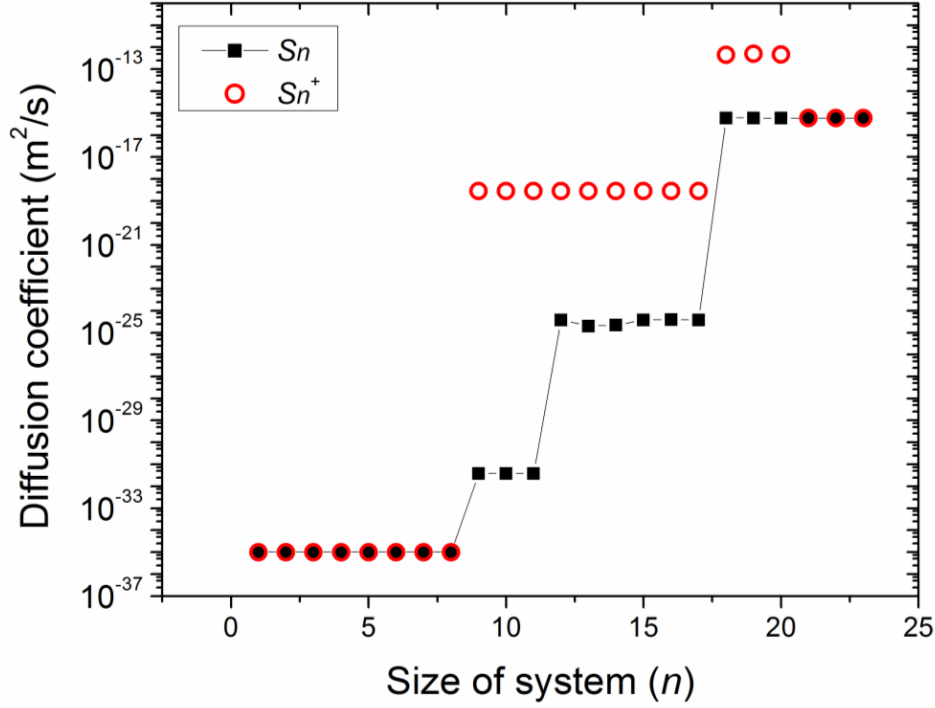

Figure 3. The diffusion coefficients calculated from kinetic Monte Carlo modeling as increasing number of Cs interstitial sites ($n$) included in the network. The Cs interstitial sites with lowest formation energies are added sequentially to the network. The $D$s are calculated with two methods, one in a system of sites and barriers associated with the $n$ points ($S_n$, black) and the other in a system of all points where additional barriers are calculated by the KRA method ($S_n^+$, red).

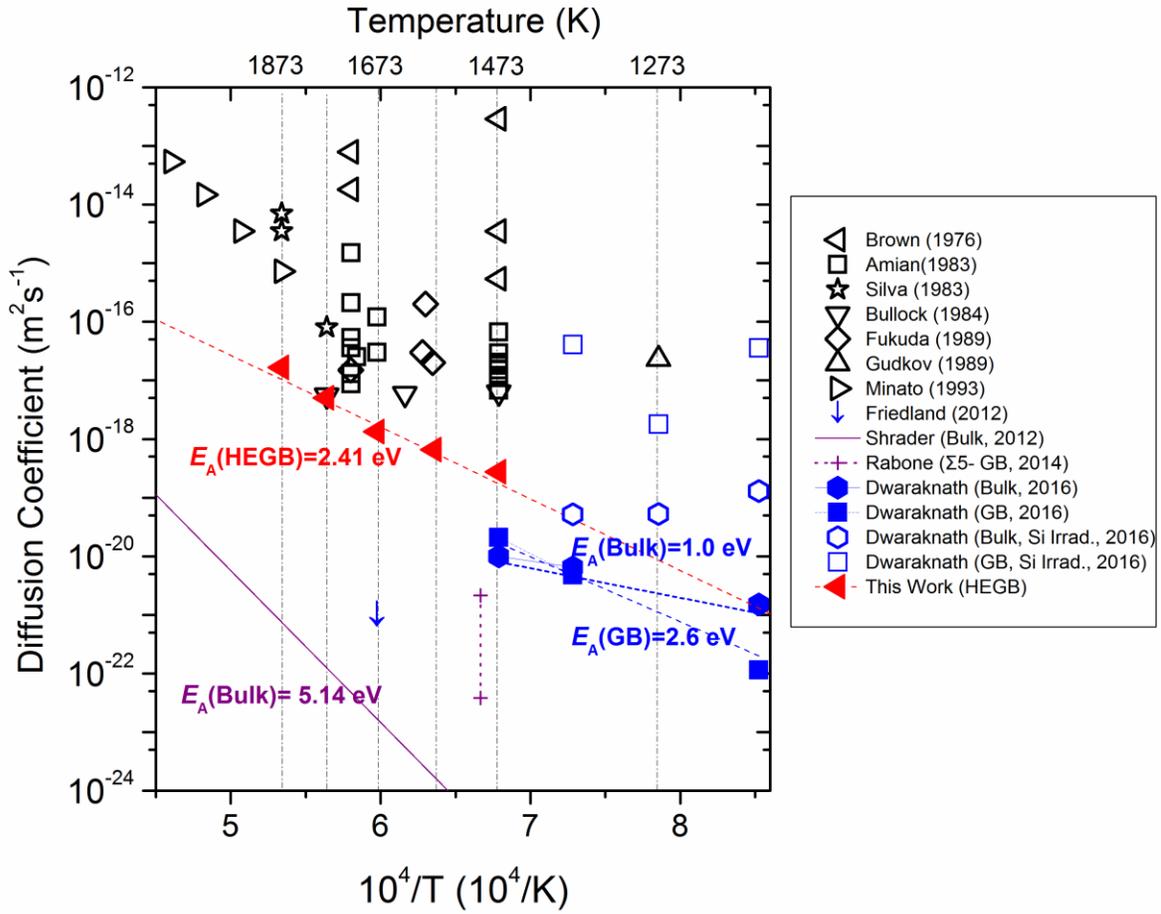

Figure 4. Summary of the temperature dependence of Cs diffusion coefficients from this work and literature. The Arrhenius fit for Cs diffusion in a HEGB from this work is shown with red dashed line. Open symbols are measurement from irradiated TRISO particles and filled symbols represents reported values from non-irradiated SiC from both surrogate experimental and computational studies. Ref [7,75-80] are integral release data from irradiated TRISO particles. Ref [13,14] are the upper bounds for $D$ for Cs in crystalline 3C-SiC and $\sum$5-GB from computational study. The Arrhenius fit for both bulk and GB Cs diffusion from ion implantation studies [68] are shown with blue dashed lines, and the fit from an *ab-initio* study [13] is shown with a purple line. All the values for Arrhenius fittings can be found in Table 1. Note the downward arrow is an indication that the value is upper limit (i.e. the estimation of $D$ is less than this value).

Table 1. A summary of reported diffusion coefficients for Cs in SiC in the form of an Arrhenius relation (Eq.(3)) when available. The temperature range shown gives the range of temperature values used to fit the Arrhenius relation for *D*.

| Reference | Temp.(°C) | $D_0 (m^2 s^{-1})$ | $E_A$ (eV) |
|---|---|---|---|
| *Integral Release* | | | |
| Allelein (1980) [81] | 1000–1600 | $1.8 \times 10^{-11}$ | 1.82 |
| Amian (1983) [7] | 1200–1450 | $3.5 \times 10^{-9}$ | 2.45 |
| Myers (1984) [82] | 700–1200 | $6.7 \times 10^{-14}$ | 1.10 |
| Myers (1984) [82] | 1600–2700 | $1.1 \times 10^{-4}$ | 4.53 |
| Bullock (1984) [78] | 1200–1500 | - | - |
| Ogawa (1985) [83] | 1300–1500 | $2.8 \times 10^{-4}$ | 4.35 |
| Moormann (1897) [84] | 1550–1900 | $2.4 \times 10^{-2}$ | 5.00 |
| Fukuda (1989) [76] | 1300–1450 | $6.8 \times 10^{-12}$ | 1.83 |
| Gudkov (1989) [79] | 1000 | - | - |
| Minato (1991) [77] | 1600–1900 | $2.5 \times 10^{-2}$ | 5.21 |
| Verfondern (1991) [3,85] | 800–1400 | $5.5 \times 10^{-14}$ | 1.30 |
| Verfondern (1991) [3,85] | 1500–2100 | $1.6 \times 10^{-2}$ | 5.33 |
| *Ion Implantation* | | | |
| Friedland [86] [a,b] | 1000–1500 | - | - |
| Dwaraknath (Bulk) [68] | 900–1200 | $2.1 \times 10^{-17}$ | 1.0 |
| Dwaraknath (GB) [68] | 900–1200 | $5.6 \times 10^{-13}$ [e] | 2.6 |
| Dwaraknath (Bulk) [72] | 900-1100 | $1.7 \times 10^{-21}$ | -0.4 |
| Dwaraknath (GB) [72] | 900-1100 | - | - |
| *Simulations* | | | |
| Shrader, Bulk [13] [c] | - | $5.1 \times 10^{-8}$ | 5.14 |
| Rabone, Σ5-GB [14] [d] | 1227 | - | - |
| This work, HEGB | 1200–1600 | $(3.75 \pm 3.15) \times 10^{-11}$ | $2.41 \pm 0.26$ |

[a] Cs diffusion in single crystal SiC, if marked

[b] upper limit of diffusion coefficient as no detectable Cs diffusion was observed

[c] by transition state theory

[d] by density functional theory molecular dynamics

[e] evaluated at T=1100°C

Table 2. The DFT formation energies ($E_f$) of Cs interstitial ($Cs_I$), and substitutionals on C ($Cs_C$) and Si ($Cs_{Si}$) lattices, under Si-rich condition. The $E_f$ in sc-SiC, and mean $E_f$ in a-SiC are summarized. For a-SiC, mean and standard deviation (s.t.d.) of $E_f$ from 15 samples are shown.

|  | sc-SiC, $E_f$ | a-SiC, mean $E_f$ | a-SiC, s.t.d. $E_f$ |
|---|---|---|---|
| $Cs_I$ | 23.46 | 0.41 | 3.40 |
| $Cs_C$ | 12.50 | 2.33 | 3.59 |
| $Cs_{Si}$ | 12.71 | 8.45 | 3.31 |